\documentclass{jac}
\usepackage{amsmath,amssymb,graphicx}

\begin{document}
\title
  {1D Effectively Closed Subshifts and 2D Tilings}
\author[lab1]{B. Durand}{Bruno Durand}
\author[lab2]{A. Romashchenko}{Andrei Romashchenko}
\author[lab2]{A. Shen}{Alexander Shen}
\address[lab1]{LIF, CNRS \&  Aix--Marseille Universit\'e}
\email{Bruno.Durand@lif.univ-mrs.fr}
\urladdr{http://www.lif.univ-mrs.fr/\~\relax bdurand}
\address[lab2]{LIF, CNRS \& Aix--Marseille Universit\'e,
on leave from IITP RAS}
\email{{Andrei.Romashchenko,Alexander.Shen}@lif.univ-mrs.fr}
%\date{}
\thanks{The paper was supported  part by NAFIT ANR-08-EMER-008-01 
grant, part by EMC ANR-09-BLAN-0164-01.}

\begin{abstract}\noindent
Michael Hochman showed that every 1D effectively closed subshift
can be simulated by a 3D subshift of finite type and asked whether
the same can be done in 2D. It turned out that the answer is positive
and necessary tools were already developed in tilings theory.

We discuss two alternative approaches: first, developed by N.~Aubrun
and M.~Sablik, goes back to Leonid Levin; the second one, developed
by the authors, goes back to Peter Gacs.
\end{abstract}

\maketitle

\section{Simulation}

Let $A$ be a finite alphabet and
let $F$ be an enumerable set of $A$-strings.  Consider all biinfinite
$A$-sequences (i.e., mappings of type $\mathbb{Z}\to A$) that
do not contain substrings from $F$. The set of these sequences is
effectively closed (its complement is a union of an enumerable set
of intervals in Cantor topology) and invariant under 
(left and right) shifts. Sets constructed in this way are called
\emph{effectively closed 1D subshifts.}

Effectively closed 2D subshifts are defined in a similar way; instead of
biinfinite sequences we have configurations, i.e., mappings of type 
$\mathbb{Z}^2\to A$, and instead of forbidden strings we have
forbidden patterns (rectangles filled with $A$-letters). Given the set
$F$ of forbidden patterns, we consider the set of all configurations where
no elements of $F$ appear. This set of configurations is closed under
vertical and horizontal shifts. If $F$ is enumerable, we get \emph{effectively
closed 2D subshifts}; if $F$ is finite, we get \emph{2D subshifts of finite type.}

2D subshifts of finite type are closely related to tilings. A \emph{tile} is a
square with colored sides (colors are taken from some finite set $C$). A \emph{tile
set} is a set of tiles, i.e., a subset of $C^4$, since each tile is determined by
four colors (upper, lower, left, and right). For a tile set $\tau$, we consider
all \emph{$\tau$-tilings}, i.e., the tilings of the entire plane by translated
copies of $\tau$-tiles with matching colors.

Tilings can be considered as a special case of 2D subshifts of finite type. 
Indeed, subshift is governed by local rules (forbidden pattern say what
is not allowed according to these rules). In tilings the rules are extremely local:
they say that the neighbor tiles should have matching colors, i.e., $1\times 2$
rectangles where the colors do not match, are forbidden.

So every tile set determines a subshift of finite type (the alphabet is a tile
set). The reverse statement is also true if we allow the extension of an alphabet.
(It is natural since we have to simulate any local rule by a more restricted
class of matching rules.)  Formally, for every alphabet $A$ and subshift $S$
of finite type we can find: 

\textbullet\  a set of colors $C$;

\textbullet\ a tile set $\tau\subset C^4$;

\textbullet\ a mapping $d\colon\tau\to A$

\noindent
such that every $\tau$-tiling after applying $d$ to each tile becomes an
element of $S$ and every element of $S$ can be obtained in this way from some
$\tau$-tiling. Such a correspondence between tilings and subshifts of
finite type works in any dimension: $k$-dimensional tilings correspond to $k$-dimensional subshifts of finite type (modulo the alphabet extension).

Now we want to compare
subshifts in different dimensions. Let $S$ be a 1D subshift. We can make a 2D
subshift from it by copying each letter vertically. It is easy to see that
an effectively closed 1D subshift becomes an effectively closed 2D subshift
(we use rules that guarantee the vertical propagation, i.e., require that vertical
neighbors should have the same letter, and the rules of the original 1D
subshift in horizontal direction). This 2D shift, denoted by $\bar S$, 
is not of finite type, if the original
1D shift was not of finite type. However, $\bar S$ is \emph{sofic}, i.e.,
is a projection by a subshift
of finite type in extended alphabet:

\begin{theorem}
\label{hochman}
	For every effectively closed 1D subshift $S$ in alphabet $A$
	there exists an alphabet $A'$, a finite $2D$ subshift  $S'$ in
	alphabet $A'$, and a mapping $d\colon A'\to A$ such that
	the image of $S'$ under $d$ \textup(applied in each 
	place\textup) is $\bar S$.
\end{theorem}

This theorem (with 3D instead of 2D, which makes it easier) was proved by
Michael Hochman~\cite{hochman} who asked whether the same is true for
2D. His motivation came from ergodic theory. 

It turned out that the tools needed to prove theorem~\ref{hochman}
for 2D tilings were already
developed in the framework of tilings theory when Hochman asked his
question. Moreover, there are two different 
sets of tools that can be used; one was used by Nathalie Aubrun and Mathieu
Sablik~\cite{sablik} (and goes back to Leonid Levin~\cite{dls}), the other
one was used in~\cite{long} (and goes back to Peter G\'acs~\cite{gacs2d}). 
In the sequel
we discuss informally how these tools work, and what are the similarities
and the differences.

\section{Tools}

Let us describe informally our problem. In 2D we have local rules that 
guarantee that each vertical line contains some letter. We need to add
some other rules to guarantee that the emerging horizontal sequence of
letters does not have substrings from some enumerable set $F$. We are
allowed to superimpose additional structure to the configuration
(by extending the alphabet: we let $A'$ be a product of $A$ and some
other finite set). Rules for this extended configuration should guarantee
that its base belongs to $\bar S$.

So we need to run a computation that generates $F$ and some process
that compares generated elements with substrings in the horizontal
sequence. It is well known (since the first papers 
of Wang~\cite{wang61,wang62} where the
notion of a tile set was introduced) that tile sets can simulate computation
easily: indeed, a time-space diagram of a Turing machine (or a cellular
automaton) obeys local rules that guarantee that computation is performed
correctly when started. The problem is to initiate the computation: 
there is no special point in the plane where the computation can be
started, so we need to ``break the translational symmetry'' somehow.

This problem was solved by Berger~\cite{berger} 
who proved that there exists an
aperiodic tile set, i.e., a tile set $\tau$ such that $\tau$-tilings exist
but all are aperiodic. (A tiling is \emph{periodic} if there is a non-zero
translation that does not change it. One can show that if a tile set
has a periodic tiling then it has a $2$-periodic tiling where
some finite block is repeated horizontally and vertically.) 
Berger used a complicated 
multi-level construction that was later simplified in different ways
by Robinson~\cite{robinson} and others. The simplification made clear that
Berger's construction is essentially based on self-similarity: 
any tiling can be divided into blocks that behave like individual tiles.
(In the original construction this similarity was obscured by
some irregularities; the cleaned versions could be found in~\cite{ollinger}
or~\cite{mathint}.)

This self-similarity creates some kind of a skeleton that can be used
to initiate computations. However, the problem is that we necessarily
initiate them in many different places, and these ``geometrically parallel''
computation should be organized to achieve some goal. Berger used them
to prove the undecidability of the domino problem (to determine whether
a given tile set has at least one tiling); for that purpose it is enough to
initiate multiple copies of the same computation: all are limited in 
time and space, but among them
there are computations of arbitrary length. For that
we split the plane into different zones used for different
computations. It is possible to find such an arrangement;
in each zone the standard local rules for a computation
are used but zones are not contiguous. So we need additional efforts
to transmit the information from one zone to another one. This all can be
done (with limited overlap, so the total density of information in a given
cell remains finite).

Then Hanf~\cite{hanf} and Myers~\cite{myers} 
proved that there are tile sets that admit
only non-recursive tilings (a much stronger statement
than the existence of an aperiodic tile set).
This was done by embedding a separation
problem for two inseparable enumerable sets, and for this we need
that all the parallel computations not only share the same (finite) program,
but also share the same (infinite) input. Therefore, some additional
machinery is needed to synchronize the inputs of all the computations
(each computation gets a finite part of the infinite input sequence,
but these finite parts are consistent pieces of an infinite input).

When simulating 1D effectively closed subshift, we need more:
the input is given to us externally (the contents of the vertical
lines that carry $A$-letters) and we need to check this
input against all possible forbidden substrings. This means that
we are very limited in space (and cannot distribute pieces of
input sparsely over the entire plane as before).

\subsection{Robinson-type solution}

The way to do this was developed in~\cite{dls}.  At each level
of self-similarity we have \emph{computation squares} that
are arranged in \emph{computational stripes}. Such a stripe
is infinite in vertical direction and carries an infinite computation
of a finite-space cellular automaton. (One can wonder
whether it makes sense to have an infinite computation
in a finite space. Indeed, it is not really infinite; it runs for
some time (exponential in the width of the stripe) and
then is restarted. The repeated computations are not necessarily
identical, since they interact with the other computations
which could be different.)  Each stripe performs some checks
for the part of the horizontal sequence that is near it. When
the level (and the size of the stripe) increases, the checked zone
and the time allowed for the computation increase. Working 
together, the stripes can check the horizontal sequence against
all forbidded substrings.

In~\cite{dls} this technique was used for one specific
1D effectively closed subshift with a binary alphabet
(for some fixed $\alpha<1$
we forbid all sufficiently long strings whose Kolmogorov
complexity is less than $\alpha$ times length). However,
this technique is quite general and can be used for any
1D effectively close subshift modulo some technical
problem.

This technical problem is that the underlying self-similar
structure may be ``degenerate'' in the sense that the plane
is divided in two parts that have no common ancestors.
In this case we need some additional tricks (extending the
zone of responsibility of each stripe) that were not needed for
the specific subshift of~\cite{dls}. The reason
why they were not needed: if a string of low complexity
(compared to length)
is split into two parts, one of then has low complexity, too.

So the technique of~\cite{dls} is not enough.
The final construction was discovered 
(in fact, independently from~\cite{dls})
by N.~Aubrun and M.~Sablik.

\subsection{Fixed-point solution}

There is a different way to organize the computations that uses
fixed-point self-similar tiling. The idea of a self-similar fixed-point
tile set can be explained as follows. We already know (since Wang
papers) that tiling can be used to simulate computations. This
computation, in its turn, can be used to guarantee the desired
behavior of bigger blocks, called \emph{macro-tiles}. So for a desired
behavior of macro-tiles we can construct tiling rules (i.e., tile set)
that guarantees this behavior. If, by chance, these tiling rules coincide
with the rules for macro-tiles, we get self-similarity as a consequence.

But there is a classical tool to get this coincidence intentionally,
not by chance: the Kleene fixed-point construction. It was used
by Kleene in the recursion theory and later by von Neumann to
construct self-reproducing automata. Usually it is illustrated as
follows: for every program $p$ (in fact, for every string $p$)
there exists a program $p'$ that prints the text of $p$. Kleene's
theorem guarantees that one can find $p$ such that $p'$ is
equivalent to $p$, i.e., the program $p$ prints its own 
text. The same trick (though not just the statement of Kleene's recursion
theorem) can be used for
2D computations. This was done first by G\'acs~\cite{gacs2d} in a complicated
setting (error-correction in 2D computations); we use the same idea in a 
much simpler environment. For each tile set $\tau$ one can construct
a set $\tau'$ of tiles that force macro-tiles to behave like $\tau$-tiles;
Kleene's trick can then be used to make $\tau$ isomorphic to $\tau'$.
This construction is explained in~\cite{short}.

Then some additional structure can be superimposed with this
self-similar skeleton (by adding some other computations); Kleene's 
trick can still be used to achieve self-similarity (in some extended
sense).

This construction is rather flexible and can be applied to different 
problems, see~\cite{long}. The differences and similiarities between
two constructions are summarized in the following comparison
table.

\subsection{Comparison table}

\bigskip
\begin{center}
\begin{tabular}{p{0.3\textwidth}p{0.3\textwidth}p{0.3\textwidth}}

Problem & Solution 1 & Solution 2 \\
     \hline
Breaking the symmetry &
Use (modified) Berger--Robinson self-similar construction
where self-similarity is guaranteed by geometric arguments &
Use fixed-point self-similar construction, where self-similiarity
is a byproduct of some computational structure \\
     \hline
Placing the computations &
Computations of different levels are all performed
          ``on the ground'', by
     	individual cells, and the plane is divided into regions
	allocated to each level &
Computations of different levels are performed
          at different levels of hierarchy: high level computations
          deal not with individual tiles but with macro-tiles\\
      \hline     
Arranging arbitrarily long computations &
Computations are infinite in the vertical
          direction but finite in horizontal direction,
          each computation performs a space-bounded
          check of some part of the horizontal sequence;
          the bound increases with the level &
Computations are finite in both direction; each computation
          performs a time-bounded check of some part of
          the horizontal sequence; the bound increases
          with the level\\
\hline
Bringing the bits of the horizontal sequence to the computation&
          Recursively from lower levels; the bits are synchronized explicitly
          ``on the ground''&
          Recursively from lower levels; each level checks whether the
          bits at the next level are recorded correctly\\ 
         \hline
Dealing with degenerate case of the self-similar pattern &
Using overlapping zones of responsibility &   
Using overlapping zones of responsibility\\
\hline
Error resistance &
Not clear (we first need some error-resistant underlying geometric
construction) &
Adding redundancy at each level\\

\end{tabular}
\end{center}


\begin{thebibliography}{9}

\bibitem{sablik}
N.~Aubrun, M.~Sablik, 
Simulation of recursively enumerable subshifts by two dimensional SFT and a generalization. Preprint, available from M.~Sablik's home page.

\bibitem{berger}
R.~Berger, The undecidability of the domino problem. \emph{Memoirs
of the AMS}, v.~66 (1966).

\bibitem{mathint}
B.~Durand, L.~Levin, A.~Shen,
Local rules and global order, or aperiodic tilings,
\emph{The Mathematical Intelligencer}, v.~27 (2005), no.~1,
p.~64--68.

\bibitem{dls}
B.~Durand, L.~Levin, A.~Shen, Complex Tilings.
\emph{J. Symbolic Logic}, \textbf{73} (2), 593--613, 2008.

\bibitem{short}
B.~Durand, A.~Romashchenko, A.~Shen, Fixed Point and Aperiodic Tilings.  
\emph{Proc. 12th International Conference of Developments in Language
Theory. Kyoto, Japan, 2008}, p.~537--548. 

\bibitem{long}
B.~Durand, A.~Romashchenko, A.~Shen, Fixed-point tile sets and their applications.  CoRR abs/0910.2415, 2009. {\tt http://arxiv.org/abs/0910.2415}

\bibitem{gurevich} 
B.~Durand, A.~Romashchenko, A.~Shen, 
Effective closed subshifts in 1D can be implemented in 2D.
\emph{Fields of Logic and Computation}, Lecture Notes in Computer
Science, v.~6300 (2010), p.~208--226.

\bibitem{gacs2d}
P.~G\'acs, Reliable Computation with Cellular Automata. J. Comput. Syst. Sci. \textbf{32}(1), 15--78, 1986.

\bibitem{hanf}
W.~Hanf, Nonrecursive tilings of the plane, i, \emph{Journal of Symbolic
Logic}, v.~39 (1974), no.~2, p.~283--285.

\bibitem{hochman}
M. Hochman, On the dynamic and recursive properties of multidimensional symbolic systems. \emph{Inventiones mathematicae}, \textbf{176}, 131--167 (2009).

\bibitem{myers}
D.~Myers, Nonrecursive tilings of the plane, ii, \emph{Journal of Symbolic
Logic}, v.~39 (1974), no.~2, p.~286--294.

\bibitem{ollinger}
N.~Ollinger, Two-by-two Substitution Systems and the Undecidability of the
Domino Problem, \emph{Computability in Europe, 2008} (CiE'2008),
Lecture Notes in Computer Science, v.~5028, p.~476--485.

\bibitem{robinson}
R.~Robinson, Undecidability and nonperiodicity for tilings of the plane,
\emph{Inventiones Mathematicae}, v.~12 (1971), p.~177--209.

\bibitem{wang61}
H.~Wang, Proving theorems by pattern recognition, II,
\emph{Bell System Technical Journal}, v.~40 (1961), p.~1--41.

\bibitem{wang62}
H.~Wang, Dominoes and the $\forall\exists\forall$ case of the 
decision problem. \emph{Proceedings of the Symposium on
Mathematical Theory of Automata}, Brooklyn Polytechnic Institute,
New York, 1962, p.~23--55.

\end{thebibliography}
\end{document}